\documentclass{emulateapj}
\pdfoutput=1 
\usepackage{hyperref}
\usepackage{amsmath,amstext}
\usepackage[T1]{fontenc}
\usepackage{apjfonts} 
\usepackage[figure,figure*]{hypcap}
\usepackage{url}
\usepackage{booktabs,threeparttable}
\usepackage{color}

\usepackage[normalem]{ulem}
\shorttitle{Searching for a superburst oscillation signal in regular thermonuclear bursts}
\shortauthors{van der Wateren, Watts \& Ootes}

\begin{document}

\title{A search for the $835\,\text{Hz}$ superburst oscillation signal in the regular thermonuclear bursts of 4U 1636-536}
\author{Emma van der Wateren\altaffilmark{1,}\altaffilmark{2}, Anna L. Watts\altaffilmark{3}, Laura S. Ootes\altaffilmark{3}}
\affil{$^{1}$ASTRON, Netherlands Institute for Radio Astronomy, Oude Hoogeveensedijk 4, 7991 PD, Dwingeloo, The Netherlands}
\affil{$^{2}$Department of Astrophysics, Radboud University Nijmegen, PO Box 9010, 6500 GL Nijmegen, The Netherlands}
\affil{$^{3}$Anton Pannekoek Institute for Astronomy, University of
  Amsterdam, Postbus 94249, 1090 GE Amsterdam, the Netherlands}
\altaffiltext{1}{Email: wateren@astron.nl}

\begin{abstract}
Burst oscillations are brightness asymmetries that develop in the burning ocean during thermonuclear bursts on accreting neutron stars.  They have been observed during H/He-triggered (Type I) bursts and Carbon-triggered superbursts.  The mechanism responsible is not unknown, but the dominant burst oscillation frequency is typically within a few Hz of the spin frequency, where this is independently known.  One of the best-studied burst oscillation sources, 4U 1636-536, has oscillations at $581\,\text{Hz}$ in both its regular Type I bursts and in one superburst. Recently however, Strohmayer \& Mahmoodifar reported the discovery of an additional signal at a higher frequency, $835\,\text{Hz}$, during the superburst. This higher frequency is consistent with the predictions for several types of global ocean mode, one of the possible burst oscillation mechanisms. If this is the case then the same physical mechanism may operate in the normal Type I bursts of this source. In this paper we report a stacked search for periodic signals in the regular Type I bursts: we found no significant signal at the higher frequency, with upper limits for the single trial root mean square (rms) fractional amplitude of 0.57(6)\%. Our analysis did however reveal that the dominant $581\,\text{Hz}$ burst oscillation signal is present at a weak level even in the sample of bursts where it cannot be detected in individual bursts. This indicates that any cutoff in the burst oscillation mechanism occurs below the detection threshold of existing X-ray telescopes.
\end{abstract}


\maketitle

\section{Introduction}

The layer of hydrogen and/or helium that builds up on the surface of an accreting neutron star can, under certain conditions, explode.  This occurs because the nuclear burning processes that take place as the matter is compressed are highly temperature-sensitive, and prone to runaway.  The resulting thermonuclear bursts manifest as Type I X-ray bursts; substantial increases in X-ray luminosity that last typically $\sim 10-100\,\text{s}$  \citep[for reviews see][]{Bildsten98b,Strohmayer06,Parikh13}.   On rare occasions more energetic bursts known as superbursts, with durations of a few hours, are also observed \citep{Cornelisse00,Strohmayer02b,Keek12}.  These are triggered by explosively unstable burning of a deep carbon layer \citep{Cumming01,Cooper09}, which is itself generated by the burning of lighter elements \citep{Stevens14}.  

Some (although not all) Type I bursts show burst oscillations, anomalously bright patches on the burning surface that give rise to pulsations in X-ray luminosity as the star rotates \citep{Strohmayer96b,Galloway08}.  For stars whose spin frequency is known independently (via the presence of accretion-powered pulsations), the burst oscillation frequency is at most a few Hz from the spin frequency.  This indicates that the bright patch is near-stationary in the rotating frame of the star.  The mechanism responsible for burst oscillations remains unknown, but possibilities include flame confinement or the development of global modes of oscillation in the burning ocean for a review of burst oscillation properties and models see \citet{Watts12}.   

Detecting burst oscillations requires high time-resolution X-ray data: obtaining sufficient photons with the instruments available to date has necessitated pointed observations.  Observing regular Type I bursts during scheduled pointed observations is relatively straightforward, since they typically recur every few hours.  For superbursts this is far more challenging, since they occur at most every $1-2$ years.  Only two superbursts have ever been observed, by pure chance, during pointed observations with a high time resolution instrument, the {\it Rossi X-ray Timing Explorer} ({\it RXTE}) in both cases.  In one case the high time resolution data of the onset of the burst were lost and data were captured for only the decaying part of the burst \citep{Strohmayer02b}. There have been no reports of burst oscillation detections in this data. During the other superburst, from 4U 1636-536, burst oscillations were detected at $582\,\text{Hz}$ \citep{Strohmayer02a}.  This is very close to the $580-581\,\text{Hz}$ burst oscillations seen in the regular Type I bursts of this source \citep{Strohmayer98a,Muno02a}.   The $582\,\text{Hz}$ superburst oscillations (SBOs) were detectable for nearly 800s near the peak of the superburst, and showed a frequency drift compatible with Doppler shifts due to orbital motion of the neutron star around the centre of mass of the binary system. This suggested that the underlying frequency was very stable, in contrast to the burst oscillations seen in the regular bursts of this source, which show drifts of $1-2 \,\text{Hz}$ \citep{Muno02a}. While the oscillation frequency in individual bursts tends to drift, the evidence shows a long-term stability of the overall distribution of these oscillation frequencies \citep{Strohmayer98b,Giles02}.

Recently, \citet{Strohmayer14b} re-visited the 4U 1636-536 superburst.  Using {\it RXTE} data from the long-lived $582\,\text{Hz}$ SBO to determine the best-fit orbital ephemeris, they corrected the photon arrival times to remove orbital phase shifts. This enabled a more sensitive search for signals that would otherwise be smeared across several frequency bins by orbital Doppler shifts, over the entire duration of the superburst.  It resulted in the detection of a signal at $835\,\text{Hz}$, with a probability (determined using Monte Carlo simulations, and after accounting for numbers of trials) of $1.5 \times 10^{-4}$ of arising solely as the result of noise.   The rms fractional amplitude of the $835\,\text{Hz}$ SBO, at ($0.13 \pm 0.03$) \%, is a factor of a few lower than the fractional amplitude of the $582\,\text{Hz}$ SBO over its detection interval \citep{Strohmayer02a}. 

Assuming that the spin frequency of the star is $582\,\text{Hz}$ (the main SBO and burst oscillation frequency of this source) \citet{Strohmayer14b} noted that an $835 \,\text{Hz}$ signal would be consistent with predictions for several types of global mode that may exist in neutron star oceans \citep{McDermott87, Strohmayer96a, Bildsten96, Heyl04, Piro05b, Strohmayer14a}.  At present there is however too much modelling uncertainty to decide between the possible mode types:  core r-modes (driven by the Coriolis force, rendered visible via coupling to the surface ocean layers), rotationally-modified ocean g-modes (driven by buoyancy or compositional discontinuities), or an ocean-crust interfacial mode associated with the discontinuity at this depth.    

One piece of information that may help theorists to identify the nature of the $835 \,\text{Hz}$ SBO is whether it can be excited to detectable levels in the regular Type I bursts of this source.  Global mode frequencies are set by stellar properties such as core/crust/ocean structure/composition, overall mass and radius, temperature, rotation rate, and perhaps magnetic field effects in the upper ocean.  Some of these factors vary sufficiently slowly that they may be considered to be constant on timescales of years (the time period for which we have observations of bursts and superbursts).  The core, in particular, is unlikely to vary significantly on the timescales of interest, and hence core mode frequencies should be the same in bursts and superbursts (if the modes can be excited, see below).  Other conditions will differ: superbursts, for example, are more energetic and ignite deeper in the ocean, hence heat the lower layers more effectively.  So if the modes are driven by oceanic processes, and the frequencies set by temperature in the deeper layers, one might not expect to find the same frequency in bursts and superbursts.  Mode excitation conditions also differ between bursts and superbursts:  superburst ignition, unlike Type I burst ignition, is likely to involve the generation of shock waves \citep{Weinberg06b,Keek12}.  Superburst ignition takes place at greater depth (at densities $\sim 10^8\,\text{g\:cm}^{-3}$ as compared to $\sim 10^5-10^6\,\text{g\:cm}^{-3}$ for Type I bursts).  Superbursts are also much more energetic and last for longer, which could be important if the modes require time to grow to detectable amplitudes.  

In this paper we search the full sample of Type I bursts from 4U 1636-536 obtained over the lifetime of the {\it RXTE} for any evidence of the $835\,\text{Hz}$ SBO frequency.  In addition to searching individual bursts we also perform a stacked search, combining data from multiple bursts, to increase sensitivity to weak signals.  No significant signal at or near $835\,\text{Hz}$ was found in either individual or stacked searches.  However a stacked search of bursts that {\bf individually} show no sign of the main $581\,\text{Hz}$ burst oscillation did reveal a significant peak at this frequency.  The implications of both of these findings are discussed in Section \ref{disc}.

\section{Analysis}
\label{analysis}

\subsection{Data selection}

{\it RXTE}, which launched on December 30 1995 and operated until January 5 2012, was until recently the only X-ray telescope with the time resolution and sensitivity to detect burst oscillations from 4U 1636-536. This has now changed with launch of Astrosat \citep{Singh14} and NICER \citep{Arzoumanian14}, but the {\it RXTE} archive remains the largest for this source. {\it RXTE}'s primary instrument, the Proportional Counter Array (PCA), consisted of five xenon-filled proportional counters sensitive to photons with energies of $2-60$ keV \citep{Jahoda96}.  During its lifetime {\it RXTE} carried out multiple observations of 4U 1636-536, recording a total of 381 Type I bursts \citep[see][and the MINBAR database\footnote{The MINBAR database, maintained by Dr D. Galloway, can be found at \url{http://burst.sci.monash.edu/minbar}.}, which extends this earlier catalogue to the end of {\it RXTE}'s lifetime.]{Galloway08}.  

We then discarded a number of bursts from our sample, following the same procedure as \citet{Ootes17} to ensure that our burst samples are consistent with that paper.  We eliminated the following:  

\begin{itemize}
\item{14 bursts that were marked with one of the following flags in either the \textit{RXTE} or MINBAR database: e, f, or g  \citep{Galloway08}. These flags indicate: e) Very faint bursts, for which only the burst peak could be observed, and no other parameters could be determined. f) Bursts that are either very faint or bursts for which there were problems with the background subtractions, such that no spectral fit of the burst could be obtained. g) Bursts that were only partly observed, resulting in an unconfirmed burst.}
\item{28 bursts with a minimum background-subtracted burst count of below 5000 photons within the first 16 seconds of the burst.  This was too few for the timing analysis conducted in \citet{Ootes17}, which is relevant to this paper since we use the detection or non-detection of $581\,\text{Hz}$ burst oscillations from that study to group the bursts.}
\end{itemize}
The other exclusion criteria detailed in \citet{Ootes17} (which addressed a larger sample of sources) did not apply to any of the bursts from 4U 1636-536.  In total, 42 bursts were eliminated from the sample, leaving 339 bursts in total for analysis.  The remaining burst data consisted of 125$\mu$s time resolution event mode data from the PCA. 

\subsection{Methodology}

We search for periodic signals using power spectra.  Photon arrival times are first binned to form a lightcurve $x_k(t)$ (counts per time bin $t_k$, where $k = 1...N$) at time resolution $\Delta t = 1/4096\,\text{s}$.  The Nyquist frequency $f_\mathrm{Ny}$, set by the time resolution, is $2048\,\text{Hz}$.  The power spectrum $P_j$ at the Fourier frequencies $\nu_j = j/T$ ($j = 0, 2, ..., N/2$ where $\nu_{N/2} = f_\mathrm{Ny}$ and $T$ is the total duration of the lightcurve), using the standard Leahy normalisation \citep{Leahy83}, is then given by 

\begin{equation*}
P_j = \frac{2}{N_\gamma} \left[\left(\sum^N_{k=1} x_k \cos   2\pi \nu_j t_k\right)^2 + \left(\sum^N_{k=1} x_k \sin 2\pi
    \nu_jt_k\right)^2\right],
\end{equation*}
where $N_\gamma = \sum^N_{k=1} x_k $ is the total number of photons.  In the absence of any deterministic signal, the powers should be distributed as $\chi^2$ with 2 degrees of freedom (d.o.f.), and we can use the properties of this distribution to assess the significance of any candidate high power, taking into account the number of trials (e.g. number of bursts and frequency bins searched).  

In this paper we average (stack) power spectra from many different bursts to maximize sensitivity to weak signals, and average powers from $W$ neighbouring frequency bins (effectively rebinning in frequency resolution), to maximize sensitivity to drifting signals.  Averaging modifies the theoretical distribution of noise powers to $\chi^2$ with 2$n$ degrees of freedom, where $n$ is the number of power spectra averaged \citep{vanderKlis89}. If stacking power spectra from $M$ burst segments of different duration, and hence native frequency resolution, so that $W$ is different for each burst, 

\begin{equation*}
n = \sum_{i=1}^M W_i
\end{equation*}
In practice, the distribution of powers in the absence of a periodic signal never precisely matches the theoretical distribution, particularly at low frequencies where the overall rise and fall of the burst envelope becomes significant. However, for the high frequencies that we study in this paper the theoretical distributions are very close to being correct \citep[see the discussion in][]{Watts12}.   

\subsection{Analysis}

We carry out various different searches, grouping the sample according to whether or not bursts show the $581\,\text{Hz}$ burst oscillation frequency and the phase of the burst (rise or decay).   Since our sample is the same, we used the criteria set out in \citet{Ootes17} \citep[similar to the criteria used in earlier work by][]{Muno04,Galloway08} to determine whether or not a burst shows the $581\,\text{Hz}$ oscillation, and we refer the reader to Section 3.2.4 of that paper for more details.  On the basis of these criteria we divide the bursts into two groups:  Sample 1 (82 bursts) with individual detections of the $581 \,\text{Hz}$ burst oscillations; and Sample 2 (257 bursts) without.  

The start of the burst $t_s$ is defined, just as in \citet{Ootes17}, as the point where the count rate first exceeds 1.5 times the pre-burst count rate.  Peak time $t_p$ is defined as the time at which the maximum count rate is reached, and end time $t_e$ is defined as 1.5 times the pre-burst count rate. A time resolution of $0.25\;\text{s}$ was used to determine these time frames.  In order to stack power spectra they need to have the same frequency resolution, which means that the duration of each burst phase separately needs to be an even multiple of the shortest length \citep[see the discussion in][who conducted a stacked search for burst oscillations from EXO 0748-676]{Villarreal04}.  The average duration for the 4U 1636-536 burst sample is $\approx 25\;\text{s}$, so when making power spectra for the full bursts we make intervals from 4 to $40\;\text{s}$ starting at $t_s$.  We analyze rise and decay portions separately. The average duration of the rise ($t_p - t_s$) for the 4U 1636-536 burst sample is $\approx 4\;\text{s}$, resulting in rise intervals of duration $\tau_r$ from 1 to $8\;\text{s}$. We define the rising phase for all bursts as being photons that arrive in the time window $t_s$ to $t_s + \tau_r$. For the decay of the bursts, we take intervals with duration $\tau_d$ in the range 4 to 32$\;\text{s}$ and define the decay phase as photons in the window $t_p$ to $t_p + \tau_d$. 

We also need to consider the issue of frequency resolution.  In their superburst analysis, \citet{Strohmayer14b} took a long stretch of data, for which the native frequency resolution was high, and corrected for orbital frequency shifts.  There was no binning to reduce frequency resolution.  In analyzing the regular bursts we have much shorter stretches of data, resulting in a frequency resolution of at best $0.25\,\text{Hz}$, determined by the shortest length. The first question we must consider is the effect of orbital Doppler shifts, since there is no reliable ephemeris covering the entire burst data set (which spans 16 years). We can estimate the size of the effects, however, using the ephemeris of \citet{Strohmayer02a}.  During a single burst, the shift in frequency of an $835\,\text{Hz}$ signal would be at most $\sim 10^{-5}\,\text{Hz}$, far below the achievable frequency resolution.  However, we are going to combine bursts occurring at different phases in the binary orbit: over the course of the orbit, a baseline frequency of $835\,\text{Hz}$ could shift by $\pm 0.38\,\text{Hz}$.  The best way to deal with this, in the absence of a good ephemeris, is to rebin to reduce the frequency resolution (by averaging neighbouring frequency bins).  Note that we do not barycenter the data: the maximum drift that might arise from {\it RXTE}'s motion around the Earth, and Earth's orbit around the Sun, is smaller, at $\sim \pm 0.1\,\text{Hz}$.   The other reason for considering reduced frequency resolution is to account for drifts in frequency.  Although both $582\,\text{Hz}$ and the $835\,\text{Hz}$ SBO frequencies were stable (the latter appears confined to 1 high resolution frequency bin), the $581\,\text{Hz}$ frequency seen in the regular bursts drifts by typically $1-2\,\text{Hz}$ \citep{Muno02a}.  Since we do not know the origin of the $835\,\text{Hz}$ signal, it is unclear whether we might expect this to drift in the regular bursts \citep[in response, for example, to the rapidly evolving thermal state of the ocean, see][]{Piro05b}, or to remain stable (if it originates in the core or crust). We must allow for the possibility of drifting.

We compute averaged power spectra for Sample 1, Sample 2 and Sample 1 and 2 combined for three different time frames: the full bursts, the rise of the bursts and the decay of the bursts. We analyze the power spectra in two frequency resolutions: $0.25\,\text{Hz}$ and $1\,\text{Hz}$ for the full burst and the decay, and $1 \,\text{Hz}$ and $5 \,\text{Hz}$ for the rise. In all eighteen averaged power spectra we search for a signal by determining the maximum power $P_\mathrm{max}$ in all frequency bins $\pm 5 \,\text{Hz}$ around the signal frequencies of $581\,\text{Hz}$ and $835\,\text{Hz}$.
We determine whether a measured $P_\mathrm{max}$ results from a significant signal by computing the probability of $P_\mathrm{max}$ or higher arising from noise alone. For this we use the following cumulative distribution function for averaged powers:
\begin{equation}
\label{pdet}
f_n(P:0) = 1-\exp(Pn/2) \sum^{n-1}_{m=0}\frac{1}{m!}\left(\frac{Pn}{2}\right)^m.
\end{equation}
\citep[derived following the prescription described in][but using the appropriate normalisation]{Groth75}. Here $f_n(P:0)$ is the probability that a measured power $P$ is between 0 and $P$ when there is no signal present, for a power spectrum with $n$ averaged spectra. From this distribution we obtain the probability $1-f$ that a measured power would exceed $P$ in the absence of a signal. We also use this distribution to compute detection thresholds $P_\mathrm{det}$. The detection threshold is the minimum power for which the probability of it being measured in the absence of a signal is less than $3\sigma$ ($1.350 \times 10^{-3}$ for a one sided test), taking into account the number of trials. 

Parts of the {\it RXTE} data show deadtime, resulting in a small reduction in the Poisson noise and a mean value for the noise power less than 2 (see \citet{Bilous18} for more on this). If we take this effect into account, the detection threshold would be slightly lower than the values we get from Equation \ref{pdet}. We can estimate this effect by fitting $\chi^2$ distributions to the noise powers of our power spectrum, where the powers around the signal frequencies of $581 \,\text{Hz}$ and $835 \,\text{Hz}$ are removed. From these distributions we again compute the detection thresholds $P_\mathrm{det}$; we find these to be 2 -- 3\% lower than the ones computed with Equation \ref{pdet}.

For each $P_\mathrm{max}$ we compute the average signal power $P_\mathrm{S}$, which is the power giving rise to the averaged measured power $P_\mathrm{max}$. We do one of two things. (1) For a detected signal we calculate the most likely signal power. Since an averaged power spectrum with high $n$ will show an average noise of 2 \citep{vanderKlis89}, for $P_\mathrm{max} \geq P_\mathrm{det}$, $P_\mathrm{S} \approx P_\mathrm{max} - 2$.  (2) For a non-detection ($P_\mathrm{max} < P_\mathrm{det}$) we compute the upper limit of the signal power which would give rise to $P_\mathrm{max}$ or higher 99.7\% of the time. For this we use the cumulative distribution function $f_n(P:P_\mathrm{S})$, which gives the probability that an averaged power $P$ is measured between 0 and $P$ given an averaged signal power $P_\mathrm{S}$:
\begin{equation*}
f_n(P:P_\mathrm{S}) = 1-\exp{[-(P+P_\mathrm{S})n/2]} \sum^\infty_{k=0} \sum ^{k+n-1}_{m=0} \frac{P_\mathrm{S}^kP^mn^{m+k}}{m!k!2^{m+k}}.
\end{equation*}
\citep[as above, derived following the prescription described in][but using the appropriate normalisation]{Groth75}.  The upper limits for $P_\mathrm{S}$ are computed without considering numbers of trials.\\
With $P_\mathrm{S}$ we determine the root mean square (rms) fractional amplitude $r$ of the power spectra:
\begin{equation*}
r = \left(\frac{P_\mathrm{S}}{N_\gamma}\right)^{\frac{1}{2}}\left(\frac{N_\gamma}{N_\gamma - N_\mathrm{B}}\right),
\end{equation*}
where $N_\mathrm{\gamma}$ is the average number of photons per power spectrum and $N_\mathrm{B}$ is the average number of background photons, estimated by using pre-burst data. 

The uncertainty in the fractional amplitude is computed by assuming Poisson noise for $N_\mathrm{\gamma}$ and $N_\mathrm{B}$. We estimate the error on $P_\mathrm{S}$ by fitting $\chi^2$ distributions to the noise powers and calculating the 1$\sigma$ standard deviation.
\section{Results}
We compute power spectra from each time interval, for Sample 1 and Sample 2. First we compute spectra for each burst individually. The individual bursts from Sample 1 show a signal at $581\,\text{Hz}$. This signal is not seen in the individual bursts in Sample 2. No significant signal is found at $835\,\text{Hz}$ for any of the individual bursts in Sample 1 or 2. 

\begin{figure}[th!]
\includegraphics[width=9.0cm]{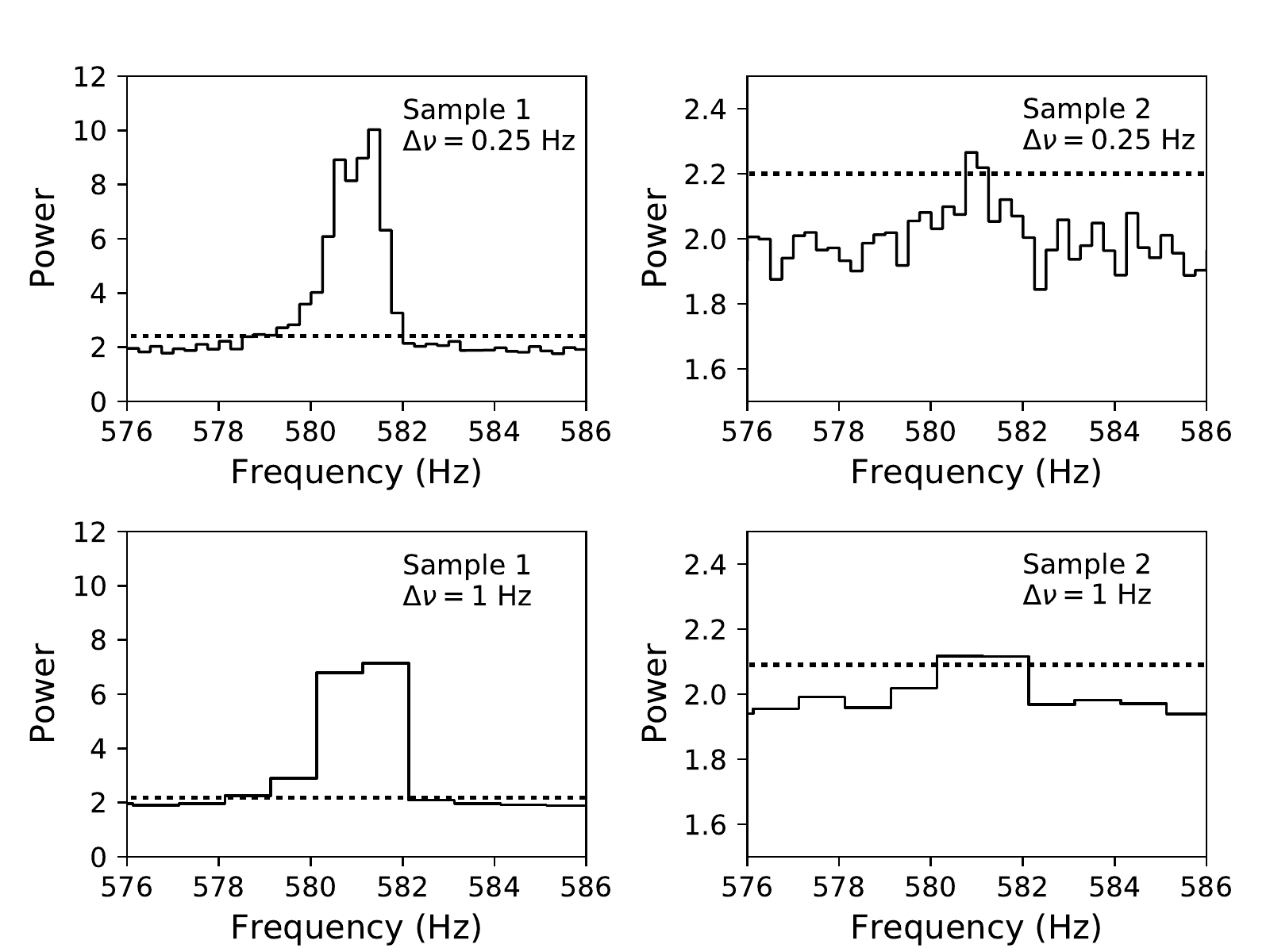} 
\caption{\small Stacked power spectra of the complete bursts from Sample 1 (Left) and Sample 2 (Right), at different frequency resolutions (Upper - $0.25\,\text{Hz}$, Lower - $1\,\text{Hz}$. The dotted line marks the detection threshold.)
}
\label{spectra}
\end{figure}

Next we stack the power spectra from the individual bursts  and look for $P_\mathrm{max}$ in a frequency window of $\pm 5\,\text{Hz}$ around both $581\,\text{Hz}$ and $835\,\text{Hz}$.  Table \ref{pscomp} summarizes the results for the various different stacked power spectra that were computed, including the significance of any detections and upper limits in the event of non-detections.  Figure \ref{spectra} shows the results when taking power spectra of the full bursts.  The power spectra taken from the complete bursts show a significant signal around $581\,\text{Hz}$ for both Sample 1 and 2. There is no significant signal around $835\,\text{Hz}$ in either sample or the samples combined, with an upper limit of $P_\mathrm{S}$ = 0.51 and an rms fractional amplitude of 0.57(6)\% in Sample 1 and $P_\mathrm{S}$ = 0.25 and an rms fractional amplitude of 0.58(7)\% in Sample 2.

As can be seen in the upper panels in Figure \ref{spectra} the signal is very strong around $581\,\text{Hz}$ for Sample 1, but can also be detected, just above the threshold, in the complete bursts from Sample 2.  No signal was found around $835\,\text{Hz}$. We reduce the frequency resolution and again search for a signal around $835\,\text{Hz}$. Again we do not detect the $835\,\text{Hz}$ SBO frequency in any of the power spectra. 

Taking into account any possible deadtime in the data would slightly lower the detection threshold. This may be relevant for any weak signals in the data. We check if any of the power spectra show a previously undetected signal, if we apply the detection thresholds from fitting $\chi^2$ distributions to the noise, as outlined in the previous section. With these slightly lower detection thresholds we still do not find a significant signal around $835\,\text{Hz}$. Two of the non-detections around $581\,\text{Hz}$ would be deemed significant in this case, as marked in Table \ref{pscomp}.

\begin{table*}
    \centering  
\begin{threeparttable}
\caption{Stacked power spectra computed}  

\begin{tabular}{lllllllllllll} 
\hline
\hline
\multicolumn{6}{r}{$581 \;\text{Hz}$}&
  \multicolumn{4}{r}{$835 \;\text{Hz}$\ }&\\
  
   \cmidrule(lr){5-8}    \cmidrule(lr){9-12}
  
Sample & $\Delta \nu$ & $n$ & $P_\mathrm{det}$ & $P_\mathrm{max}$\tnote{a} & $f$ & $P_\mathrm{S}$\tnote{b} & $r\tnote{b}\:\:\:\:(\%)$& $P_\mathrm{max}$\tnote{a} & $f$ & $P_\mathrm{S}$\tnote{b}& $r\tnote{b}\:\:\:\:(\%)$ 

\\
\hline
\\
Full burst\\
1 + 2 & $0.25\,\text{Hz}$ & 2172  & 2.18 &   3.98 &  $1.6 \times 10^{-287}$ & 1.98 & 1.49(2)& \textit{2.06} &  $8.2 \times 10^{-2}$ & 0.19 &  0.46(6)  \\
1 + 2 & $1.0\,\text{Hz}$ & 8688  &  2.08 &  3.33 & $2.0 \times 10^{-588}$ & 1.33 & 2.44(6)&  \textit{1.97} & 0.92  & 0.03 & 0.4(3) \\
1 & $0.25\,\text{Hz}$ & 429  & 2.41 & 10.02 &  $6.3 \times 10^{-450}$ &  8.02 & 2.27(2)& \textit{2.19} &  $2.7 \times 10^{-2}$ & 0.51 & 0.57(6) \\
1 & $1.0\,\text{Hz}$  & 1716 & 2.18 & 7.15 & $9.5 \times 10^{-973}$ & 5.15 & 3.64(6)& \textit{2.05} &  $0.15$ & 0.19 & 0.7(2)  \\
2 & $0.25\,\text{Hz}$  & 1743  & 2.20 & 2.27 &  $3.2 \times 10^{-8}$ & 0.27 & 0.60(6)& \textit{2.10} & $2.0 \times 10^{-2}$& 0.25& 0.58(7) \\
2 & $1.0\,\text{Hz}$ & 6972  & 2.09 & 2.12 & $4.4 \times 10^{-7}$ & 0.12 & 0.8(2) & \textit{2.00} & 0.50 & 0.07 & 0.6(3)\\
\hline
\\
Rise\\
1 + 2 & $1.0\,\text{Hz}$ & 1364   & 2.20 &  4.06 & $2.0 \times 10^{-193}$ & 2.06 & 2.70(6)& \textit{2.03} & 0.29 & 0.19 & 0.8(2) \\
1 + 2 &$5.0\,\text{Hz}$  & 6820  & 2.08 & 2.63 & $1.8 \times 10^{-124}$ & 0.63 & 3.3(2)& \textit{1.95} & 0.98 & 0.02 & 0.6(1) \\
1 & $1.0\,\text{Hz}$ & 166   & 2.61 & 9.98 & $1.6 \times 10^{-174}$ & 7.98 & 3.45(5)& \textit{2.33} & $2.0 \times 10^{-2}$ & 0.89 & 1.2(1) \\
1 & $5.0\,\text{Hz}$ & 830 & 2.23 & 4.37 & $4.6 \times 10^{-148}$ & 2.37 & 4.2(2) & \textit{2.08} & $0.13$ & 0.29 & 1.5(5)\\
2 & $1.0\,\text{Hz}$ & 1198  & 2.22 & 2.23 & $6.0 \times 10^{-5}$ & 0.23 & 1.0(2) & \textit{1.96} & 0.75 & 0.13 & 0.7(2) \\
2 & $5.0\,\text{Hz}$  & 5990  & 2.08 & \textit{2.07}\tnote{c} & $3.07 \times 10^{-3}$ & 0.15 &1.8(4) & \textit{1.95} & 0.97 & 0.02 &0.7(1) \\
\hline
\\
Decay\\
1 + 2 & $0.25\,\text{Hz}$  & 1799 & 2.19 & 4.11 & $2.7 \times 10^{-264}$ & 2.11 & 1.57(2) & \textit{2.07} &$7.0 \times 10^{-2}$ & 0.21& 0.41(8)\\
1 + 2 & $1.0\,\text{Hz}$ & 7196 & 2.09 &  3.39 & $8.6 \times 10^{-526}$ & 1.39  & 2.54(6)& \textit{1.98} & 0.80 & 0.05 & 1.0(1) \\
1 & $0.25\,\text{Hz}$  & 377 & 2.44 & 10.57 & $4.9 \times 10^{-432}$ & 8.57& 2.39(2) & \textit{2.22} & $1.9 \times 10^{-2}$ & 0.56& 0.61(6)\\
1 & $1.0\,\text{Hz}$ & 1512 & 2.19 & 7.47 & $2.5 \times 10^{-931}$ & 5.47 & 3.83(6)& \textit{2.07} & $8.8 \times 10^{-2}$ & 0.23 & 0.8(2) \\
2 &$0.25\,\text{Hz}$ & 1421 & 2.22 & \textit{2.17} & $8.8 \times 10^{-4}$ & 0.29 & 0.65(7) & \textit{2.11} & $2.0 \times 10^{-2}$ & 0.27& 0.63(7) \\
2 & $1.0\,\text{Hz}$ & 5684 & 2.10 & \textit{2.09}\tnote{c} & $4.1 \times 10^{-4}$ & 0.17 & 1.0(1) & \textit{2.01} & 0.35 & 0.09 & 0.7(3) \\
\hline
\end{tabular}
    \begin{tablenotes}
      \small
            \item NOTE -- We show the results for the various different stacked power spectra that were computed from Sample 1, Sample 2 and Sample 1 + 2 combined. Here $\Delta \nu$ is the frequency resolution of the power spectra, $n$ is the amount of added power spectra, $P_\mathrm{det}$ is the detection threshold for a significant signal, taking into account number of trials, $P_\mathrm{max}$ is the maximum power, $f$ is the probability that the measured power $P_\mathrm{max}$ is between 0 and $P_\mathrm{max}$ when there is no signal present, $P_\mathrm{S}$ is the signal power and $r$ the rms fractional amplitude, with the $\pm 1 \:\sigma$ error in the last digit in brackets.\\
      \item[a] We determine $P_\mathrm{max}$ in a frequency window of $\pm 5\,\text{Hz}$ around the signal frequencies of $581$ and $835\,\text{Hz}$. Numbers written in italic signify non-detections.
  \item[b] For the non-detections the quoted $P_\mathrm{S}$ and $r$ are upper limits for a single trial.
  \item[c] These non-detections would be deemed significant if we used the fitted $\chi^2$ distribution for the noise powers, in an effort to account for dead-time (see text).
    \end{tablenotes}
\label{pscomp}
\end{threeparttable}
\end{table*}

\section{Conclusion and discussion}  
\label{disc}
The $835\,\text{Hz}$ SBO frequency cannot be found in either the individual Type I bursts or either of the stacked burst samples. If it is excited in the Type I bursts, then its amplitude must remain very low: we find an upper limit on the rms fractional amplitude of 0.57(6)\% for the complete bursts of Sample 1 (those in which the dominant 581 Hz frequency is detected) and 0.58(7)\% for the bursts of Sample 2 (those for which the 581 Hz signal is not detected in any individual burst). 

The question is then why the $835\,\text{Hz}$ signal can be detected in the superburst but not in the Type I bursts. One possibility is simply that the mechanism that generates the $835\,\text{Hz}$ signal in the superburst does not operate in the normal bursts: perhaps it is associated with burning at greater depths in the accreted layers of the star. An alternative is that the mechanism takes longer than a typical burst duration to be excited to detectable levels:  Type I bursts last for only $\sim 10-100$ seconds, while the superburst displaying the $835\,\text{Hz}$ signal lasted several hours \citep{Strohmayer14b}.  This poses a new constraint on theory: any model that purports to explain the SBO oscillation must not operate under normal Type I burst conditions, except at very weak levels.

Our analysis also revealed something interesting about the dominant $581\,\text{Hz}$ burst oscillations.   In 82 of the bursts from 4U 1636-536, this oscillation signal can be detected in individual bursts; but in 257 bursts the oscillation signal is not detectable in individual bursts. However a stacked search of this latter group of bursts does result in a significant signal at  $581\,\text{Hz}$. From this we can conclude that the signal still exists at a weak level, below the detection threshold for individual bursts. This indicates that the cutoff in the mechanism responsible for exciting the burst oscillations occurs below the detection threshold of existing instruments. A telescope with a larger collecting area, such as the proposed Enhanced X-ray Timing and Polarimetry mission \citep[eXTP,][]{Zhang19,intZand19} or the Spectroscopic Time-Resolving Observatory for Broadband Energy X-rays \citep[STROBE-X,][]{Ray18}, will be able to explore this discovery space and answer the question of whether the mechanism has a cut-off at low amplitude. We have also shown that stacking bursts can be a useful method for detecting weak signals, despite any possible frequency drifting. In this case however, we knew where to look for the signal. If we would have searched the entire spectrum, resulting in up to 8192 number of trials, some signals in Sample 2 would not have been considered significant.\\

\acknowledgments

EvdW undertook this analysis for her Bachelors Project Thesis at the University of Amsterdam.  ALW acknowledges support from ERC Starting Grant No. 639217 CSINEUTRONSTAR.  LSO acknowledges support from NWO Top Grant Module 1 (PI Rudy Wijnands).  
\\
\\

\bibliographystyle{yahapj}
\bibliography{sbosc}

\end{document}